\documentclass[epj]{svjour}
\usepackage{amssymb,amsmath}
\usepackage{graphics}
\usepackage{widetext}
\begin{document}
\title{Creation of electron-positron plasma with superstrong laser field}
\author{N.B.~Narozhny \and A.M.~Fedotov}
\institute{National Research Nuclear University MEPHI, 115409 Moscow, Russia}
\date{Received: date / Revised version: date}
\abstract{We present a short review of recent progress in studying QED effects of interaction of ultra-relativistic laser pulses with vacuum and $e^-e^+$ plasma. The development of laser technologies promises very rapid growth of laser intensities in close future already. Two exawatt class facilities (ELI and XCELS, Russia) in Europe are already in the planning stage. Realization of these projects will make available a laser of intensity $\sim 10^{26}$W/cm$^2$ or even higher. Therefore, discussion of nonlinear optical effects in vacuum are becoming urgent for experimentalists and are currently gaining much attention. We show that, in spite of the fact that the respective field strength is still essentially less than $E_S=m^2c^3/e\hbar=1.32\cdot 10^{16}$V/cm, the nonlinear vacuum effects will be accessible for observation at ELI and XCELS facilities. The most promissory for observation is the effect of pair creation by laser pulse in vacuum. It is shown, that at intensities $\sim 5\cdot 10^{25}$W/cm$^2$, creation even of a single pair is accompanied by development of an avalanchelike QED cascade. There exists an important distinctive feature of the laser-induced cascades, as compared with the air showers arising due to primary cosmic ray entering the atmosphere. In our case the laser field plays not only the role of a target (similar to a nucleus in the case of air showers). It is responsible also for acceleration of slow particles. It is shown that the effect of pair creation imposes a natural limit for attainable laser intensity. Apparently, the field strength  $E\sim E_S$ is not accessible for pair creating electromagnetic field at all.
} 
\maketitle

\section{Introduction}
\label{intro}

The invention of the CPA (Chirped Pulse Amplification) technique \cite{Mourou} in the second half of 80s and its subsequent development has resulted in creation of petawatt laser facilities generating short pulses of coherent optical radiation with peak intensities up to $10^{22}$W/cm$^2$ \cite{Yanovsky}. Currently, two European projects, ELI (Extreme Light Infrastructure \cite{ELI}) and XCELS (Exawatt Center for Extreme Light Studies \cite{XCELS}), both aiming at generation of femtosecond laser pulses with intensity more than $10^{24}$W/cm$^2$, and in the long term -- up to $10^{26}$W/cm$^2$, are in the planning stage. This will open novel possibilities for investigation of  effects of nonlinear interaction of electromagnetic radiation with matter (for details, see the reviews \cite{rev2012}). In 2011, a special network called IZEST (International Center for Zetta-Exawatt Science and Technology) was established \cite{IZEST} in order to stimulate exchange of ideas and coordination of activities for different groups worldwide involved in the projects in the field of fundamental and high-energy physics with high intensity lasers.

One of the most challenging research trends is experimental investigation of non-linear QED effects in the presence of very strong electromagnetic fields. Dozens of theoretical works on Intense Field QED (IFQED) have been published starting from early 1960s. The probabilities of basic IFQED processes, such as emission and absorption of a photon by an electron, pair photoproduction and annihilation of a pair into a single photon, photon splitting in the field of a plane monochromatic electromagnetic wave and in a constant field, etc., have been calculated. It was predicted theoretically that polarized by the strong field QED vacuum acquires characteristics of a non-linear optical media, thus giving rise to a number of new optical effects, e.g., birefringence and dichroism of vacuum \cite{Baier67,Adler71}, photon splitting \cite{Adler71,Birula71}, Cherenkov radiation in vacuum \cite{Erber66}, mutual and self-focusing in vacuum \cite{Rozanov98}, light-by-light scattering \cite{diPiazza05,Lundstrom06}, harmonic generation in vacuum \cite{Kaplan00,diPiazza05,Fedotov07}, etc. A comprehensive overview of the work performed prior to 1985, as well as a detailed list of references can be found in \cite{NR,BKS}. 

So far, we have the only one experimental test of IFQED, namely the famous E144 experiment performed at SLAC in 1996-97. Scattering of ultrarelativistic electrons with the energy $46.6$Gev by laser pulses of the intensity $10^{18}$W/cm$^2$ (nonlinear Compton effect) \cite{Bula} and electron-positron pair production by backscatterd hard photons have been observed \cite{Burke}. The results of the experiments proved to be in quantitative agreement with the theory, see \cite{Bamber} for the details. As it was mentioned above, the intensities of modern laser facilities have increased already by four orders of magnitude, and are planned to be increased further by 2-4 orders more in the nearest future. This opens  unique opportunities for experimental investigation of the already observed IFQED effects at a new intensity level, as well as of completely new, hitherto unexplored experimentally, nonlinear vacuum effects. 

In this paper, we focus on a brief review of some recently published fundamental theoretical results related to the IFQED studies with the forthcoming laser facilities.

\section{Pair creation by laser field in vacuum}
\label{sec:pair_creation}

Perhaps, the most interesting among the non-linear IFQED effects is the electron-positron pairs creation by a strong classical electromagnetic field in vacuum. Field strength for this process is scaled by the so-called critical QED electric field strength $E_{cr}=m^2c^3/e\hbar=1.32\times 10^{16}$V/cm first introduced by F. Sauter \cite{Saut}, see \cite{Schw}. An electron gains the energy equal to $mc^2$ across its Compton length $l_C=\hbar/mc=3.86\cdot 10^{-11}$cm under the action of such field. For a long time it was mistakenly believed that the probability of the pair creation effect was solely determined by the Schwinger exponent $\exp(-\pi E_{cr}/E)$, and thus was exponentialy small at $E<E_{cr}$. However, the Schwinger exponent nominally determines the probability of pair creation only in the Compton 4-volume $V_CV_T=l_C^4/c$. If the peak electric field strength $E_0$ is small compared with $E_S$, $E_0\ll E_{cr}$, the probability of pair creation in a Compton 4-volume remains exponentially supressed. However, if the field strength $E_0$ is created in 4-volume $VT$ which is much greater than the Compton one, then the probability of pair creation acquires a large pre-exponential factor with the magnitude of the order of the ratio $ VT/V_CT_C $. For an optical laser pulse with a wavelength $\lambda=1\mu$m and duration of $10$fs, focused to a diffraction limit, this ratio is of the order of $10^{25}$. This factor is so large that it can compensate the smallness of the Schwinger exponent at some $E_0\ll E_S$. This sort of enhancement singles out the pair creation process among other vacuum polarization effects, thus nominating it as probably the most easily observable candidate among nonlinear vacuum IFQED processes.

Our method for calculating the probability of pair creation by the laser field \cite{pairs1} is based on the fact that the typical formation length $l_f$ and time $T_f$ for this process for fields close to critical are of the order of Compton length and time respectively \cite{AI}. Since for laser radiation in optical range the relation $l_f\sim l_C\ll\lambda$ holds, any such field can be locally considered as a constant and homogeneous with respect to the process of pair creation. This means that the mean number of produced pairs can be calculated with the following formula \cite {pairs1}
\begin{equation}\label{number}
N_{e^+e^-}=\frac{e^2E_S^2}{4\pi^2}\int dV\int_ {-\infty}^{\infty} dt~\epsilon\eta \coth\left(\frac{\pi\eta}{\epsilon}\right)\exp\left(-\frac{\pi}{\epsilon}\right),
\end{equation}
where integration is performed over the whole volume occupied by the field and its duration, the invariant quantities $\epsilon=\mathcal{E}/E_S$, $\eta =\mathcal{H}/E_S$ are 
defined by
$$\mathcal{E},\mathcal{H}=\sqrt{\left(\mathcal{F}^2+\mathcal{G}^2\right)^{1/2}\pm\mathcal{F}}, \; \mathcal{F}=\frac{\vec{E}^2-\vec{H}^2}2,\; \mathcal{G}=\mathbf{\vec{E}\cdot\vec{H}},$$ 
and have the meaning of the electric and magnetic field strengths respectively in a reference frame in which they are parallel. To compute the invariants $\epsilon $ and $\eta$, we use the analytical model of a laser pulse based on an exact solution of Maxwell's equations \cite{NFo}.
\begin {figure*} [ht]
\vspace*{0.2cm} 
\resizebox{\textwidth}{!}{
\includegraphics{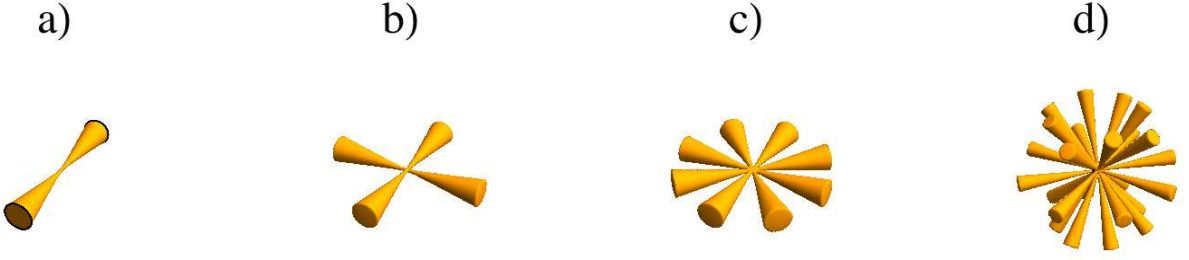}}
\caption{The concept of multi-beam technology: planer setup with a) $n=2$, b) $n=4$, c) $n=8$ colliding pulses and 3D setup d) with $n=24$ colliding pulses (as proposed in \cite{mult}).}\label{fig:multibeam}
\end{figure*}

It turned out \cite{pairs1}, that feasibility of observation of the pair creation effect for the case of collision of only two counterpropagating focused laser pulses requires intensity $\sim 10^{27}$~W/cm$^2$ which is two orders of magnitude smaller than the critical one. However, it still exceeds the capabilities of the forthcoming facilities, ELI and XCELS. Therefore it was important to find the ways for reduction of the threshold intensity  $I_{th}$ for this effect, i.e. the intensity necessary for creation of at least a single pair. To solve this problem a multi-beam technology was suggested in \cite{mult}. It was shown that for $n$ laser pulses with a fixed total power focused coherently at one point, $I_ {th}$ would decrease with increasing $n$.

Use of multi-beam setup allows to redistribute the electromagnetic field in the focal region. In the course of interference of the colliding pulses the resulting field gets there spiked spatiotemporal structure. Thus, the total 4-volume in the focal region occupied by the field is reduced, but the peak value of the field strength increases. The number of created pairs depends on the peak strength of the electric field exponentially, while on the effective focal 4-volume occupied by the field as a power. This explains the decrease of the threshold intensity. For example, two identical linearly polarized laser pulses can be always superposed in such a way that in the resulting antinodes of the arising standing wave the electric fields of the two pulses are summed but the magnetic fields compensate each other. Obviously, the most preferable multi-beam configuration is such that the central axis of the colliding pulses lie in the same plane, and the pulses are linearly polarized in the direction orthogonal to the plane of their propagation. Assuming the geometry of the experiment is organized so that all the pulses propagate in one plane and are grouped into head-on colliding couples, the magnetic fields in the common focal region will cancel each other and the electric fields will stack. Then the peak intensity of the electric field will be proportional to $\sqrt{n_ {p_1}}$, where $n_{p_1}$ is the total number of pulses. Of course, $n_{p_1}$ is limited by the aperture of the colliding pulses. In order to decrease $I_ {th}$ further, albeit with less efficiency, it is possible to add $n_{p_2}$ pulses, also grouped into counterpropagating couples with propagation axes constituting an angle $\theta$ with the plane in which the initial $n_{p_1}$ pulses are located. In this case, the resultant intensity in the focus will be proportional to $(n_{p_1}+n_{p_2}\cos\theta)/\sqrt{n_p}$, where $n_p=n_{p_1}+n_{p_2}$, \cite{mult}. 

Consider setup in which the number of colliding focused pulses reaches the value $n=24$, see Fig.~\ref{fig:multibeam}. The first eight are focused in the plane ($yz$) to form four pairs of mutually counterpropagating pulses along/opposite axes $y$ and $z$, and two	bisecting lines of the coordinate quadrants of the plane $yz$. Other eight pairs of counterpropagating pulses can be added along the lines constituting $45^0$ with the plane $(yz)$. The resulting collision geometry corresponding to 24 pulses grouped into three belts is shown in Fig.~\ref{fig:multibeam}. Note that it turns out that this setup allows one to approach very closely the theoretical limit of focusing which was discussed in Refs.~\cite{F_Lim,Gon1,Gon2}.

The results of numerical calculations for the number of pairs produced by various numbers $n$ of colliding pulses are shown Ref.~\cite{mult} in Table~\ref{tab:1}. It is assumed that all the pulses have the same duration $10$fs, wavelength $\lambda=1\mu$m and numerical aperture $\Delta=0.3$ (the latter corresponds to the size of the focal spot of the order $\lambda$). The total energy of the pulses was taken to be $10$kJ independently on $n$.
\begin{table}[h]
\label{tab:1}
\caption{The number $N_{e^+e^-}$ of electron-positron pairs created by different numbers of pulses (with numerical aperture $\Delta=0.3$ each, and the total energy $10$kJ) and the threshold value of total energy $W_{th}$.}
\begin{center}
\begin{tabular}{lll}
\hline\noalign{\smallskip}
$n$ & $N_{e^+e^-}$ at $W=10$kJ & $W_{th}$, kJ \\
\noalign{\smallskip}\hline\noalign{\smallskip}
$2$ & $9.0\times 10^{-19}$ & $40$ \\
$4$ & $3.0\times 10^{-9}$ & $20$ \\
$8$ & $1.0$ & $10$ \\
$24$ & $4.2\times 10^{6}$ & $5.1$ \\
\noalign{\smallskip}\hline
\end{tabular}
\end{center}
\end{table}
These results show that use of multi-beam technology allows significantly decrease the threshold energy for the pair creation effect as compared with the cases of a single or two colliding focused pulses. The effect can be hopefully observed at the main pillar of the ELI facility if the setup with $8$ colliding laser pulses of the total energy $10$kJ is chosen. In the arrangement with $24$ pulses the threshold energy is almost twice lower.

\section{QED cascades}
\label{sec:qed_cascades}

\begin {figure}[ht]
\vspace*{0.3cm}
\begin{center}
\resizebox{0.45\textwidth}{!}{
\includegraphics{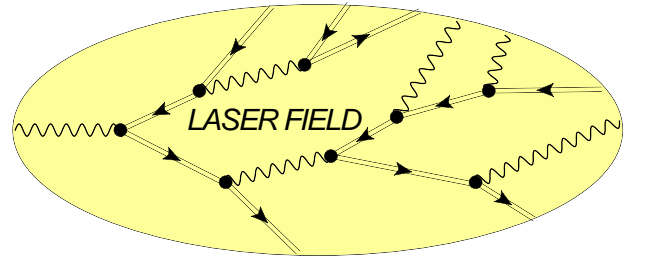}}
\end{center}
\caption{A conceptual view of QED cascade in a laser field (as an example, seeded by a hard photon arriving from the left).}\label{fig:cascade}
\end{figure}

\begin {figure}[ht]
\vspace*{0.2cm} 
\begin{center}
\resizebox{0.46\textwidth}{!}{
\includegraphics{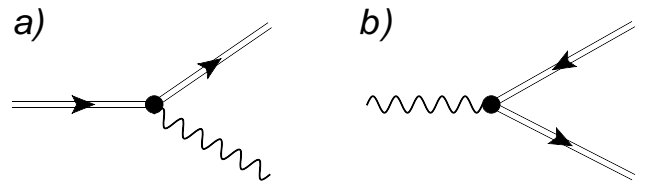}}
\end{center}
\caption{First-order QED processes in an external field: a) photon emission; b) pair photoproduction. Double solid lines directed rightwards/leftwards are electrons/positrons, dressed by the field; wavy lines are the hard photons.}\label{fig:processes}
\end{figure}

Ultra-high-intensity laser pulses, which are planned to be obtained with the forthcoming projects ELI and XCELS, open the possibility to reproduce the SLAC-like experiments at a new level \cite{Leemans}. Owing to much higher laser intensity, collisions of high-energy electrons with laser pulses will be accompanied by development of long chains (cascades) of secondary processes, instead of single events as at the SLAC experiment. These chains (Fig.~\ref{fig:cascade}) will be formed by sequential events of hard photons emissions by electrons (Fig.~\ref{fig:processes}a) and creations of
electron-positron pairs by these photons (Fig.~\ref{fig:processes}b), which may last until the charged particles totally lose their energy. In order to design experiments under such conditions it is highly desirable to simulate the laser-matter interaction processes in this novel regime.

Note that the cascade theory is well developed, especially in the context of extensive air showers (EAS) which are generated in a medium (say, planet atmosphere) by cosmic rays \cite{EAS}. Cascades in an external magnetic field are very similar and had been also well studied already \cite{Akhiez,bolg}. For the case of a laser field, the cascades development is governed \cite{Elkina2011} by the so-called cascade equations (\ref{casc1}),(\ref{casc2}), where $f_\pm(\vec{r},\vec{p}_e,t)$ and  $f_\gamma(\vec{r},\vec{p}_\gamma,t)$ are the distributions of positrons, electrons and photons respectively in the phase space, $\vec{v}_e=\vec{p}_e/\varepsilon_e$ is the velocity and $\varepsilon_e=\sqrt{p_e^2+m^2}$ is the energy of an electron or positron. $dW_{rad}(\vec{p}_e\to\vec{p}_\gamma)=w_{rad}(\vec{p}_e\to\vec{p}_\gamma)\,d^3p_\gamma$ and $dW_{cr}(\vec{p}_\gamma\to\vec{p}_e)=w_{cr}(\vec{p}_\gamma\to\vec{p}_e)\,d^3p_e$ are the differential probability rates for photon emission (Fig.~\ref{fig:processes}a) and pair photoproduction (Fig.~\ref{fig:processes}b) in the external electromagnetic field, $W_{rad}(\vec{p}_e)$ and $W_{cr}(\vec{p}_\gamma)$ are the respective total probability rates for these processes.

\begin{widetext}
\begin{subequations}
\begin{align}
\frac{\partial f_{\pm}(\vec{r},\vec{p}_e,t)}{\partial t}\pm e\left[\vec{E}(\vec{r},t)+\vec{v}_e\times\vec{H}(\vec{r},t)\right]\cdot\frac{\partial f_{\pm}(\vec{r},\vec{p}_e,t)}{\partial \vec{p}_e}
=\int\limits w_{rad}(\vec{p}_e+\vec{p}_\gamma\to \vec{p}_\gamma)f_{\pm}(\vec{r},\vec{p}_e+\vec{p}_\gamma,t)\,d^3p_\gamma\nonumber\\
-W_{rad}(\vec{p}_e) f_{\pm}(\vec{r},\vec{p}_e,t)+\int\limits  w_{cr}(\vec{p}_\gamma\to\vec{p}_e)f_{\gamma}(\vec{r},\vec{p}_\gamma,t)\,d^3p_\gamma,\label{casc1}\\
\frac{\partial f_{\gamma}(\vec{r},\vec{p}_\gamma,t)}{\partial t}=\int\limits w_{rad}(\vec{p}_e\to \vec{p}_\gamma)[f_{+}(\vec{r},\vec{p}_e,t)+f_{-}(\vec{r},\vec{p}_e,t)]\,d^3p_e-W_{cr}(\vec{p}_\gamma) f_{\gamma}(\vec{r},\vec{p}_\gamma,t),\label{casc2}
\end{align}
\end{subequations}
\end{widetext}
Note that the classical part of radiation reaction need not be included separately because it is already taken into account by the first two terms of Eq.~(\ref{casc1}), see e.g. \cite{Elkina2011,diPiazza} for the details. For the case of laser field of optical frequency ($\lambda\gg l_C$) and ultra-relativistic particles, one can use these probability rates in the approximation of a locally constant crossed (${\cal F},\,{\cal G}\approx 0$) field, which are well known \cite{NR,BKS}. Then, they are exclusively determined by the dynamical quantum parameters of participating particles 
\begin{equation}\label{chi}
\chi_{e,\gamma}=\frac{\gamma_{e,\gamma}}{E_S}\sqrt{(\vec{E}+\vec{v}_{e,\gamma}\times\vec{H})^2-(\vec{v}_{e,\gamma}\cdot\vec{E})^2},
\end{equation}
where $\gamma=\varepsilon/m$ is dimensionless energy (Lorentz factor for the charged particles), $\vec{E}(\vec{r},t)$ and $\vec{H}(\vec{r},t)$ are the local values of the electric and magnetic field. Note that $\chi$ is proportional to the product of $\gamma$ and the component of the Lorentz force, orthogonal to the momentum of a charged particle. As for the total rates, they are determined by the value of $\chi$ of the incoming particle, e.g.
\begin{equation}\label{W}
W_{e,\gamma}\sim \frac{\alpha m^2}{\varepsilon_{e,\gamma}}\chi_{e,\gamma}^{2/3},\quad \chi_{e,\gamma}\gg 1.
\end{equation}
In the classical limit $\chi_\gamma\lesssim 1$ pair photoproduction is exponentially suppressed.

Cascade pair production in a media or magnetic field requires high energy of the initial seed particle and lasts until the secondary particles totally lose their kinetic energy. The multiplicity of such a cascade is always limited. For instance, in magnetic field it can be estimated simply as $N_{e^+e^-}\sim \chi_{in}$, where $\chi_{in}$ is the quantum dynamical parameter of the seed particle, compare \cite{Heitler}. Cascades in a laser field are very different. In the regions where ${\cal F}>0$ the laser field is capable of acceleration of charged particles \cite{BK,MK}, thus restoring their energy, or the value of the quantum dynamical parameter. As it was first noted in \cite{BK}, the laser field can play a dual role. In addition to being a target for high energy particles, it is a continuously working particle accelerator. Therefore, the multiplicity of QED cascades can be limited by the duration of the laser pulse rather than by the value of $\chi_{in}$ under some conditions, and the number of created secondary particles can become very large \cite{MK}. In this regard, QED cascades resemble electron avalanches occurring due to impact ionization in dielectric-filled trench used for electrical isolation of semiconductor devices \cite{sem}.

Simple estimates allowing to gain some insight into the physical mechanism of self-sustaining regime of cascade pair production were suggested in Ref.~\cite{MK}. It is useful to ignore for a moment the effect of the magnetic field and consider a model of a planar uniformly rotating electric field $\vec{E}(t)=\{E_0\cos{\Omega t},E_0\sin{\Omega t}\}$ (such field is very similar to the field in the anti-nodes of a circularly polarized monochromatic standing wave). According to the equation of motion $\dot{\vec{p}}_e=e\vec{E}(t)$, the momentum of an electron initially at rest primarily grows linearly, 
\begin{equation}\label{energy}
\varepsilon_e\sim p_e\sim eE_0 \delta t,
\end{equation}
but retards with respect to the rotating direction of the field. So that, the angle between them also grows linearly with $t$, $\theta_e\sim \Omega \delta t$. Hence, the increasing transverse, with respect to the electron momentum, component of the electric field $E_\perp\sim E_0\theta_e\sim E_0\Omega \delta t$ arises. As a result the dynamical quantum parameter starts to grow 
\begin{equation}\label{chi_est}
\chi_e\sim \gamma_e E_\perp/E_S\sim m\Omega (E_0/E_S)^2 \delta t^2,
\end{equation}
and reaches the value $\sim 1$ for $\delta t\sim t_{acc}$, where $t_{acc}=(E_S/E_0)(m\Omega)^{-1/2}$ was called the acceleration time. On the other hand, substituting (\ref{energy}) and (\ref{chi_est}) into Eq.~(\ref{W}) one can find that the mean free time $t_{e,\gamma}$, defined by $W_{e,\gamma}t_{e,\gamma}\sim 1$, is given by $t_{e,\gamma}\sim (E_S/\alpha^3E_0)^{1/4}(m\Omega)^{-1/2}$. Hence, if $E_0\gtrsim E_*=\alpha E_S\approx (1/137)E_S$, $t_{acc}\lesssim t_{e,\gamma}$. This means that the basic QED events occur with $\chi_{e,\gamma}\gtrsim 1$, or even $\chi_{e,\gamma}\gg 1$. Therefore: (i) usage of Eq.~(\ref{W}) is indeed legal, at least qualitatively; (ii) pair photoproduction is not suppressed. The latter is obviously a necessary condition for the cascade development. Multiplicity of a cascade initiated by a single seed particle at the time moment $t$ can be now estimated as $N_{e_+e^-}(t)\sim \exp(t/t_{e,\gamma})$. If one assumes that all the particles, being ultra-relativistic, vacate the focal region with almost the speed of light, then in the worst case $t\sim \Omega^{-1}$ and $t/t_{e,\gamma}\sim \sqrt{m\alpha^2/\Omega} (E_0/E_*)^{1/4}$. For an optical laser $\sqrt{m\alpha^2/\Omega}\sim 1$, therefore it turns out that the value $E_*$ is indeed the threshold strength for cascade development. It corresponds to intensity $\sim 10^{25}$W/cm$^2$. Note that according to (\ref{chi_est}), the effective expansion parameter of IFQED perturbation theory is always $\ll 1$ for all conceivable at the moment experimental setups since even for $\chi_{e,\gamma}\gg1$, according to Ref.~\cite{NB_80}, it is $\sim\alpha [\chi_{e,\gamma}(t_{e,\gamma})]^{2/3}\sim E_0/E_S\ll 1$. Hence, the QED processes of higher orders do not compete with the basic processes of the first order.

Unlike the standard theory of cascades \cite{EAS,Akhiez,bolg}, the acceleration mechanism of self-sustaining cascades essentially requires $2D$ or $3D$ treatment, which makes numerical calculations much more complicated. By now, Eqs.~(\ref{casc1}), (\ref{casc2}) were solved numerically by the Monte-Carlo method \cite{Legkov,Elkina2011,Nerush2011c,Kirk2011,Nerush2011b,Bashmakov} in combination with the particle-in-cell (PIC) scheme \cite{Nerush2011a} for the cases when it was necessary to take into account plasma effects. These simulations seem to confirm the above described qualitative picture of cascade  development. As was expected, the multiplicity primarily grows exponentially but at the moment when plasma density reaches the value $a_0n_{cr}$, and hence the plasma becomes non-transparent, rapid depletion of the laser field at the backs of colliding pulses occurs and growth of the number of pairs ceases. It is worth noting that self-sustaining cascades cannot occur for some configurations of the external field. It happens for example in the nodes of the magnetic field of a standing wave \cite{Bulanov2010} because electrons follow the direction of the field and are not deflected. A plane wave field serve the second example because the dynamical quantum parameter $\chi$ is an integral of motion in such field and thus mechanism of acceleration does not exist. However, the cascades always arise for a focused laser pulse or for any combination of several such pulses. Moreover, the threshold intensity observed in simulations was essentially lower than $10^{25}$W/cm$^2$ because of the large size of the focal region and presumably fluctuations of parameters $\chi$ and mean free times $t_{e,\gamma}$ \cite{BK,Legkov,Elkina2011,Kirk2011,Bashmakov}. But one should bear in mind that most of the authors for simplicity used in their simulations the probability rates for nonpolarized particles. However, spin effects may be of some importance and it was shown recently that not taking them into account introduces the $10\%$ error in the final results \cite{King}.

It is interesting whether impact of cascades is in general crucial for dynamics of laser-target interaction at intensities that will be attained with ELI and XCELS facilities. In Refs.~\cite{Ridgers2013,Pukhov} such interaction was simulated for moderate intensities $\sim 10^{23-24}$W/cm$^2$ for the case of normal or inclined incidence of weakly focused laser pulses onto foil targets. Contrary to what was initially anticipated, the yield of the pairs, although been observed, remained small, so that such interaction mainly results in generation of hard photons in quantitative agreement with \cite{Bulanov2012}, where pair photoproduction effect was ignored. The reason was simple. Namely, electrons and positrons were picked up and blown away towards the direction of propagation of the laser field. Such situation is known to be unfavourable for cascade development. However, the situation must be certainly different if the target is pressed by several pulses as in Fig.~\ref{fig:multibeam}, or if the laser field is tightly focused. Note that for realistic setup of laser-target interaction (e.g., for solid target) pair photoproduction on ions \cite{Liang} would seemingly exceed that one in the laser field, so that both channels must be taken into account in further simulations.

\section{Discussion}
\label{sec:discussion}

IFQED processes at the next generation of high intensity laser facilities is one of the topics which is widely discussed in literature \cite{Design2011,Leemans}. In our opinion, observation of pair production from vacuum and self-sustaining QED cascades could be among the most important goals. The issue of principal importance is weather intensity attainable with optical laser is limited, as it was deduced in \cite{MK}. Let us remind that this limitation originates from the assertion that such high intensities could be obtained seemingly only with short tightly focused laser pulses. Such fields would create pairs in vacuum, which in turn would seed the self-sustaining QED cascades catalyzing the pair creation process and thus depleting the original laser pulse. Consequently, there would be no way to overcome some threshold intensity value, which was roughly estimated to be about $3\cdot 10^{26}$W/cm$^2$. Just such intensity level was originally aimed by ELI and XCELS.

To understand physics of laser-matter interaction at the next intensity level deeper, several problems still need to be solved. Let us mention here just a few of them. First, there are some indications \cite{Nina,AMR,Zhidkov} that when radiation reaction dominates the charged particles could be trapped in the focal region and stay there much longer than it was anticipated. Second, as for now it is not enough evident what may happen at the final stage of the cascade. On the one hand, some instabilities associated with the presently employed numerical algorithms make simulations at extreme intensity level unreliable. On the other hand, at high enough plasma densities, which are typical for the final stage of the cascade, relaxation processes can come into play thermalizing the plasma and preventing further depletion of the laser field. Finally, in order to make more definite predictions, specifically on the attainability of extreme intensities, pair creation from vacuum should be systematically and precisely incorporated into the general framework of the kinetic approach, based on Eqs.~(\ref{casc1}), (\ref{casc2}). This task is still far from the completion in spite of many attempts to do it, see \cite{Gies2010} for the most recent reconsideration of the problem, and references therein. These and other associated problems are now under active investigation.

\begin{acknowledgement}
The work was supported by the Russian Fund for Basic Research (grants 11-02-12148ofi-m and 13-02-00372), the RF Ministry of Science and Education  within the Federal Program ``Scientific and scientific-pedagogical personnel of innovative Russia 2009-2013'' (agreements 14.\-A18.\-21.\-0773), and the RF President programs for support of young Russian scientists and the leading research schools (grant MD-5838.2013.2).
\end{acknowledgement}

\end{document}